\documentclass[runningheads]{llncs}
\usepackage[T1]{fontenc}
\usepackage{graphicx,verbatim}
\usepackage{hyperref} 
\usepackage{amsmath} 
\usepackage{bm} 
\usepackage{amssymb} 
\usepackage{booktabs} 
\usepackage{multirow} 

\begin{document}
\title{Spatio-Temporal Conditional Diffusion Models for Forecasting Future Multiple Sclerosis Lesion Masks Conditioned on Treatments}
\titlerunning{Spatio-Temporal Diffusion Models for MS Lesion Forecasting}
%
\author{Gian Mario Favero\thanks{Corresponding author: \email{gian.favero@mail.mcgill.ca}}\inst{1,2} \and
Ge Ya Luo\inst{2} \and
Nima Fathi\inst{1,2} \and
Justin Szeto\inst{1,2} \and \\
Douglas L. Arnold\inst{1} \and
Brennan Nichyporuk\inst{1,2} \and
Chris Pal\inst{2} \and
Tal Arbel\inst{1,2}}

%
\authorrunning{G. Favero et al.}
%
\institute{
McGill University \and
Mila – Quebec AI Institute}
\maketitle              
\begin{abstract}
Image-based personalized medicine has the potential to transform healthcare, particularly for diseases that exhibit heterogeneous progression such as Multiple Sclerosis (MS). In this work, we introduce the first treatment-aware spatio-temporal diffusion model that is able to generate \textit{future} masks demonstrating lesion evolution in MS. Our voxel-space approach incorporates multi-modal patient data, including MRI and treatment information, to forecast new and enlarging T2 (NET2) lesion masks at a future time point. Extensive experiments on a multi-centre dataset of 2131 patient 3D MRIs from randomized clinical trials for relapsing-remitting MS demonstrate that our generative model is able to accurately predict NET2 lesion masks for patients across six different treatments. Moreover, we demonstrate our model has the potential for real-world clinical applications through downstream tasks such as future lesion count and location estimation, binary lesion activity classification, and generating counterfactual future NET2 masks for several treatments with different efficacies. This work highlights the potential of causal, image-based generative models as powerful tools for advancing data-driven prognostics in MS.

\keywords{Diffusion, Multiple Sclerosis, Spatio-Temporal}
\end{abstract}
\section{Introduction}

Deep learning models for image‐based personalized medicine that predict future individual patient outcomes enable early and more informed treatment interventions before irreversible disease accrual occurs. This potential is particularly important for diseases such as Multiple Sclerosis (MS) and certain cancers, which are characterized by heterogeneous evolutions and variable treatment effects. Recent work has shown success in leveraging baseline MRI to forecast patient outcomes in MS~\cite{durso-finley22a,dursofinley2023}. However, beyond predicting numerical outcomes, a crucial question remains: What if we could also predict personalized future images depicting the appearance of future pathological structure changes (e.g., new and enlarging lesions or tumor metastases) on a patient’s brain MRI? Such visual forecasts would not only increase the trustworthiness of predictions through enhanced explainability, but would also provide deeper insights into the effects of various treatments at the individual level, thereby improving clinical decision support.

In the context of MS, lesions, visible as hyperintense regions on T2‐weighted MRI, serve as key markers for tracking disease activity. New and enlarging T2 (NET2) lesions are especially important for assessing disease progression and treatment efficacy in relapsing‐remitting MS (RRMS)~\cite{Doyle2017LesionDS,Sepahvand2020CNNDO}. However, due to their extremely small size (typically less than 0.1\% of brain tissue) and irregular distribution with lesions predominantly forming in regions such as the deep white matter~\cite{MSLesionCoeff,MSLesionCoeff2}, accurately predicting the precise locations of NET2 lesions for an individual patient is challenging. Nonetheless, in many clinical contexts, the goal is not solely to pinpoint exact locations but rather to provide clinicians with a range of plausible future images that indicate the general locations and counts of lesions under different treatments. Having this capability would represent an advance in personalized medicine and enhance clinical decision support.

To address these challenges, we turn to generative AI, particularly variational diffusion models, which have emerged as a powerful tool for medical image synthesis. Pioneering works have shown that diffusion models can generate high‐fidelity medical images via progressive denoising~\cite{BRLP,sadm,favero2025conditionaldiffusionmodelsmedical}, and subsequent refinements have enabled 3D diffusion models to preserve fine anatomical details and subtle variations~\cite{wang20243d,kim2024adaptive}. However, despite several recent advances in generating high-quality brain MR images and disease progression over time~\cite{Friedrich_2024,Khader2023,Rachmadi2025}, there are currently no generative architectures that (i) predict the appearance of new or enlarging pathological structures at future time points, (ii) perform causal inference on the resulting focal pathology maps based on real treatments, and (iii) efficiently train within a high-resolution voxel space.

In this work, we present the first treatment-aware diffusion framework that uses ControlNet to predict the evolution of NET2 lesion masks in MS. Our approach generates voxel-level predictions by restructuring 3D MRI data into pseudo-2D slabs, enabling a simple training framework free of the reliance on a separately trained VAE for image compression~\cite{BRLP,wang20243d}. By integrating baseline imaging modalities (e.g., FLAIR, T2, and gadolinium‐enhanced sequences) with treatment assignment, our model enables direct comparisons across treatment arms, while supporting multiple predictions at inference to capture variability in new disease activity. 

Extensive experiments on a multi-centre dataset of 2131 patient 3D MRIs and corresponding lesion masks from randomized clinical trials for relapsing-remitting MS demonstrate that our model accurately predicts NET2 lesion masks for patients across treatments of six different efficacies. Moreover, we show that our model has the potential for real-world downstream tasks such as future lesion count and location estimation, binary lesion activity classification, and qualitative counterfactual generation. Our findings highlight the potential of data driven approaches in personalized medicine, opening the door to future research on treatment-aware generative models and their integration into clinical workflows for neurological diseases such as MS.

\section{Methodology}

Variational diffusion models (VDMs)~\cite{kingma2023variationaldiffusionmodels} approximate complex data distributions by modeling a two-part stochastic process: a forward noising process and a learned reverse denoising process. The forward process gradually perturbs data by adding Gaussian noise:
\begin{align}
    \bm{z}_t = \alpha_\lambda\bm{x} + \sigma_\lambda\bm{\epsilon}, \quad \bm{\epsilon} \sim \mathcal{N}(0, \text{I}),
\end{align}
where $\bm{x}$ is the original data sample, $\bm{z}_t$ is the noisy latent at timestep $t$, and $\lambda$ is a parameterization of the signal-to-noise ratio (SNR) that governs the trade-off between clean signal and added noise for a given $t \in [0,1]$. The reverse process aims to reconstruct $\bm{x}$ by learning to predict and subtract away the added noise, typically through a neural network $\bm{\hat{x}}_\theta(\bm{z}_t;t)$. The training objective minimizes a weighted mean squared error between the true data and the predicted data:
\begin{align}
    \mathcal{L}_{VDM} = \mathbb{E}_{\epsilon \sim \mathcal{N}(0, \text{I}), \lambda \sim p(\lambda)} \left[ \frac{w(\lambda)}{p(\lambda)} ||\bm{x} - \bm{\hat{x}}_\theta(\bm{z}_t; t)||_2^2 \right],
\end{align}
where $p(\lambda)$ is a SNR sampling distribution and $w(\lambda)$ is a weighting function that balances learning across SNRs during training.

UNet architectures~\cite{ronneberger2015unet} are widely used as the backbone for modeling the reverse distribution in image-based diffusion networks. To incorporate additional conditioning into the UNet architecture, a ControlNet~\cite{zhang2023addingconditionalcontroltexttoimage}, creates a duplicate of the UNet contraction path to serve as an adapter within the network. This trainable copy is connected to the original blocks via zero convolutions, which are $1 \times 1$ convolution layers initialized with zero weights and bias. Over time, gradients flow through the zero convolution layers, gradually allowing the adapter learn how to incorporate image conditions into the output. During training, ControlNet follows the VDM objective with additional conditioning inputs:
\begin{align}
    \mathcal{L}_{VDM} = \mathbb{E}_{\epsilon \sim \mathcal{N}(0, \text{I}), \lambda \sim p(\lambda)} \left[ \frac{w(\lambda)}{p(\lambda)} ||\bm{x} - \bm{\hat{x}}_\theta(\bm{z}_t, \bm{c}_e, \bm{c}_f; t)||_2^2 \right],
\end{align}
where $\bm{c}_e$ and $\bm{c}_f$ correspond to conditioning embeddings and images, respectively. To encourage the model to learn useful information from both conditioning styles, $\bm{c}_e$ and $\bm{c}_f$ are randomly masked during training with independent probabilities, forcing the network to attend to both sources of context.

\begin{figure}[!htbp]
    \centering
    \includegraphics[width=\textwidth]{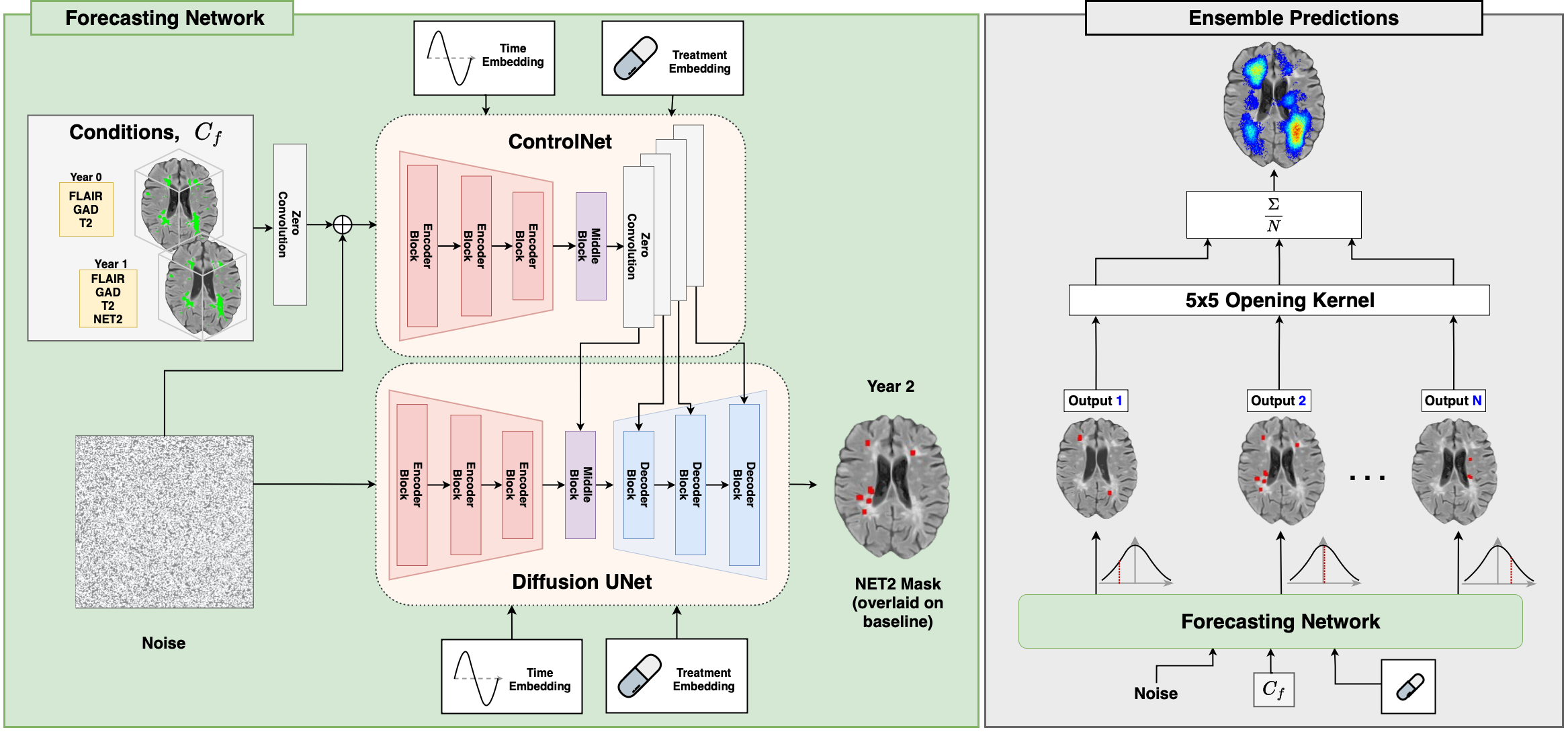}
    \caption{\textbf{Left: Model Architecture.} A treatment-conditioned diffusion UNet is first pretrained to model the distribution of future NET2. A ControlNet is then trained to enable semantic conditioning based on MRI input. \textbf{Right: Inference Pipeline.} Multiple stochastic samples are ensembled to generate a treatment-aware NET2 mask prediction for a given patient.} 
    \label{fig:architecture}
\end{figure}

The proposed method makes use of a voxel-level diffusion model that is pretrained to learn the distribution of the NET2 labels, in combination with a ControlNet, which is conditioned on patient image(s) and associated labels (e.g treatment), to predict a future, patient-specific NET2 mask. Both models are also conditioned on the treatment arm, which enables treatment specific predictions at inference. The overall architecture is depicted in Figure~\ref{fig:architecture}, and described in more detail below.

\paragraph{Stage 1 - Pre-training the diffusion model:} We first train a treatment-conditioned diffusion model to learn the distribution of future NET2 on each treatment arm. We combine the slice and channel dimensions of 3D MRI tensors ([b, c, s, h, w] $\rightarrow$ [b, c $\times$ s, h, w]), allowing the generation of 3D MRI and corresponding lesion labels with a 2D UNet~\cite{ronneberger2015unet}. Convolutional layers capture spatial relationships in the sagittal-coronal plane, while self-attention layers ensure consistency in the axial dimension. As modern accelerators offset the cost of scaling the number of channels~\cite{Hoogeboom:arXiv:2023:simpleDiffusion}, we are able to dramatically reduce the computational burden of storing 3D feature maps while allowing the use of a much higher capacity model at an equivalent batch size. 

Conditioning diffusion models on text or categorical inputs can be addressed via architectural design, such that the denoising model becomes $\hat{\bm{x}}_\theta(\bm{z}_t, \bm{c}_e; t)$ where $\bm{c}_e$ is a categorically embedded patient treatment arm assignment. Following~\cite{dhariwal2021diffusionmodelsbeatgans}, we sum the class embeddings with the time embeddings that are injected into the ResBlocks of the UNet backbone. In order to be able to use classifier-free guidance at inference~\cite{ho2022classifierfreediffusionguidance}, we simultaneously train the diffusion model to learn the unconditional NET2 distribution by randomly dropping out the treatment (using a ``null'' embedding) 10\% of the time. Moreover, we derive $p(\lambda)$ from an interpolated-shifted cosine schedule~\cite{Hoogeboom:arXiv:2023:simpleDiffusion} and use a min-SNR~\cite{hang2024efficientdiffusiontrainingminsnr} $w(\lambda)$ function.

\paragraph{Stage 2 - Training the ControlNet for predicting future NET2:} We next train the ControlNet in order to enable semantic control over generation by fine-tuning a frozen copy of a pre-trained diffusion model encoder to process conditioning images, $\bm{c}_f$, in addition to class embeddings, $\bm{c}_e$, during denoising steps, such that the model becomes $\hat{\bm{x}}_\theta(\bm{z}_t, \bm{c}_e, \bm{c}_f; t)$. Outputs of the ControlNet encoder are summed with the frozen diffusion model's feature maps at each resolution layer, integrating structural guidance without compromising the quality or diversity of the frozen diffusion model’s output. In the spirit of classifier-free guidance, we stochastically drop out both $\bm{c}_e$ and $\bm{c}_f$ during fine-tuning.

\paragraph{Inference:} At inference, the model, conditioned on patient MRI and accompanying treatment information, is used to predict the NET2 mask at a future time point via classifier-free guided sampling. Since the proposed model is inherently stochastic, each sample can produce a slightly different plausible future outcome. This uncertainty is a feature rather than a flaw: by drawing multiple samples from the model and ensembling their outputs, we obtain a more robust final prediction that reflects the model's distribution over possible NET2 outcomes.

\section{Dataset and Implementation Details}

\paragraph{Dataset:} Data are pooled from five randomized clinical trials that enrolled patients with relapsing-remitting MS (RRMS): BRAVO~\cite{Vollmer2014}, OPERA 1/2~\cite{Hauser2017}, and DEFINE~\cite{Gold2012}. Each patient sample consists of a Fluid Attenuated Inversion Recovery (FLAIR) image, lesion maps (T2 hyperintense and gadolinium-enhancing lesions), as well as relevant clinical features (treatment arm). NET2 lesion labels (identifying \textit{new} T2 hyperintense lesions) between pre-treatment (w000) and week 48 (w048), as well as between w048 and week 96 (w096) are also available. We find that the vast majority of patients only have NET2 lesion activity occurring within a 15 slice slab taken from the center of the full volume. Thus, to reduce computational burden, we crop all MRI to dimensions of (15, 256, 256) in the axial, sagittal, and coronal axes, respectively. 

In total, we construct a dataset consisting of 2131 samples belonging to the following trial arms: placebo (n=327), laquinimod (no proven efficacy, NE, n=237), avonex (low efficacy, LE, n=271), interferon beta-1a subcutaneous (mild efficacy, MiE, n=529), dimethyl fumarate (moderate efficacy, MoE, n=193), and ocrelizumab (high efficacy, HE, n=574). All MRI were acquired with 1x1x3 mm resolution at the following time points: pre-treatment (w000), one year (w048), and two years (w096).The data are split in an 80/10/10 ratio into train, validation, and test sets using stratification to preserve the treatment distribution across splits. Further statistics are provided in Table~\ref{tab:treatment-summary}.

\begin{table}[hbtp]
    \centering
    \caption{Samples, average NET2 lesion counts, and percentage of patients with NET2 lesion activity in the \textit{training set} at 1 and 2 years post-treatment. w048 values are measured relative to pre-treatment; w096 values are relative to w048.}
    \begin{tabular}{lcccccc}
        \toprule
        \textbf{Treatment} & \textbf{Samples} & \multicolumn{2}{c}{\textbf{Avg. NET2 Count}} & \multicolumn{2}{c}{\textbf{NET2 Activity (\%)}} \\
        \cmidrule(lr){3-4} \cmidrule(lr){5-6}
        & & \textbf{w048} & \textbf{w096} & \textbf{w048} & \textbf{w096} \\
        \midrule
        Placebo & 267 & 4.60 & 3.33 & 63.7 & 58.1 \\
        Laquinimod (NE) & 184 & 5.51 & 4.49 & 64.7 & 57.1 \\
        Avonex (LE) & 214 & 4.37 & 4.37 & 54.7 & 51.9 \\
        Interferon Beta-1a (MiE) & 418 & 0.63 & 1.59 & 24.9 & 33.7 \\
        Dimethyl Fumarate (MoE) & 153 & 0.60 & 1.03 & 19.6 & 26.1 \\
        Ocrelizumab (HE) & 466 & 0.07 & 0.07 & 3.20 & 1.50 \\
        \midrule
        \textbf{Total} & 1702 &  &  &  & \\
        \bottomrule
    \end{tabular}
    \label{tab:treatment-summary}
\end{table}

\paragraph{Implementation Details:} For our UNet and ControlNet backbone, we use a similar encoder path as~\cite{dhariwal2021diffusionmodelsbeatgans}, with six down/up sampling stages and self-attention at $16^2$ and $8^2$ resolutions. The diffusion UNet is pre-trained for 100k iterations on FLAIR-NET2 image pairs from w096 conditioned on trial arm. This model is then frozen and used as the initialization for the ControlNet model, which we train for 40k iterations to predict w096 NET2 labels given w000 and w048 conditioning images and the trial arm. Given the small size of the NET2 and the relatively small number of patients that are active over time, we modify the loss function to up-weight NET2 lesion areas by a factor of 10, and secondly, we train the ControlNet model on a balanced subset of the training data, ensuring an even distribution of model capacity across both active and inactive outcomes. Both models are trained on 4 A100 GPUs with a batch size of 16, and an Adam optimizer. In total, the models combine to contain 589M parameters.

\section{Experiments and Results}
We first analyze the spatial accuracy of the model's predicted NET2 lesion locations, followed by qualitative multi-inference results that highlight the model’s stochastic yet trend-consistent behavior across treatment arms. Finally, we report the model’s performance on downstream tasks that demonstrate its real-world utility. Given the lack of existing treatment-aware generative models, these experiments are evaluated against population-level statistical baselines to demonstrate relevant improvement.

\subsection{Regional Spatial Accuracy}
Accurate regional lesion localization is important for clinical interpretability. Although MS lesions play a critical role in disease assessment, they typically occupy less than 0.1\% of total brain tissue and predominantly form within the cerebral white matter, and less prominently in the cortex~\cite{Doyle2017LesionDS,Sepahvand2020CNNDO}. To ensure that our proposed model generates plausible lesions in the correct regions without explicitly relying on disease-specific priors, we compare the predicted NET2 lesion regional locations against the ground truth regional locations for patients in the test set. This comparison permits accounting for stochastic nature of the exact location of new lesional formation. The regions considered are the cerebral white matter and cerebral cortex as segmented by SynthSeg~\cite{Billot_2023}. 

\begin{table}[!htbp]
    \centering
    \renewcommand{\arraystretch}{1.2}
    \caption{Comparison of the predictive regional accuracy of our method against a Monte Carlo baseline that uses positive class proportions from population-level statistics to predict the region that NET2 lesions will appear.}
    \begin{tabular}{lcccc}
        \toprule
        & \textbf{BA}~$\uparrow$
        & \textbf{Precision}~$\uparrow$ 
        & \textbf{Recall}~$\uparrow$ 
        & \textbf{F1}~$\uparrow$ \\
        \hline
        Cerebral-WM (Proposed Method) & \textbf{0.684} & \textbf{0.537} & \textbf{0.629} & \textbf{0.579} \\
        Cerebral-WM (MC Baseline) & 0.497 & 0.321 & 0.322 & 0.320 \\
        \hline
        Cerebral-Cortex (Proposed Method) & \textbf{0.697} & \textbf{0.367} & \textbf{0.562} & \textbf{0.444} \\
        Cerebral-Cortex (MC Baseline) & 0.503 & 0.154 & 0.135 & 0.143 \\
        \bottomrule
    \end{tabular}
    \label{tab:spatial_accuracy}
\end{table}

For each brain region (cerebral white matter and cerebral cortex), we assign each patient a binary label: a ground truth label of 1 indicates the presence of at least one lesion in that region, while a label of 0 denotes its absence. Similarly, if the model predicts at least one lesion in a given region, the predicted label is set to 1, otherwise, it is set to 0. Table~\ref{tab:spatial_accuracy} compares the performance of the proposed model to a Monte Carlo baseline. For each region, the baseline predicts lesion presence by sampling from a Bernoulli distribution whose probability parameter is set to the empirical lesion prevalence for patients in the training set based on regions provided by SynthSeg segmentation~\cite{Billot_2023}—approximately 33\% for cerebral white matter and 10\% for cerebral cortex. Our model outperforms this baseline, indicating its ability to predict the spatial distribution of NET2 lesions beyond what is achievable using population-level statistics alone. Note that no other baseline exists for this problem.

\subsection{Exploring Ensembles of Multiple Inferences} 

\begin{figure}[!htbp]
    \centering    
    \includegraphics[width=\textwidth]{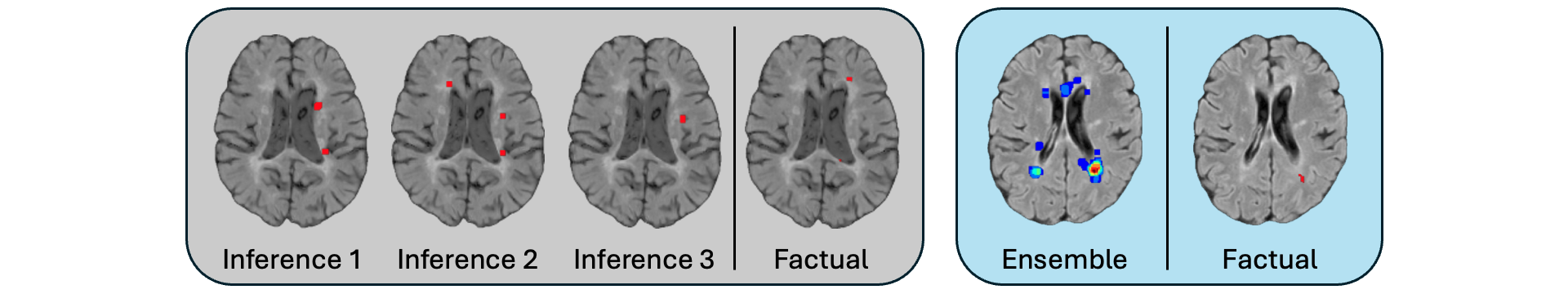}
    \caption{Left: (Placebo) Three sample inferences compared to the factual outcome. Right: (NE) A heatmap of 100 inferences for a patient and the factual outcome.}
    \label{fig:stochastic_gen}
\end{figure}

To further demonstrate the model’s ability to capture the stochastic nature of evolving MS disease activity, we perform inference several times for patients on different treatment arms. Figure~\ref{fig:stochastic_gen} presents representative examples where, for two sample treatment conditions (placebo and a treatment with no proven efficacy), central 2D slices of the generated NET2 lesion masks are overlaid on baseline MRI scans. In both cases, the patients are active with NET2 appearing in the cerebral white matter around the ventricles. For the first patient, individual inferences from the model show predicted NET2 in this region, with some variability as to the precise location. For the second patient, an ensemble of 100 inferences shows a heatmap of possible future NET2 locations. In this case, there is overlap between an area of high probability (red) and the ground truth outcome, demonstrating clinical relevance in modeling the nature of the disease.

\subsection{Relevant Downstream Tasks}
To further evaluate our proposed model, we conduct two quantitative downstream tasks from the inferred NET2 labels. Note that the model was not explicitly trained for these tasks.  

\paragraph{Binary Activity Outcome Prediction:} First, we assess the model’s ability to predict future NET2 lesion activity by classifying whether a patient will have at least one NET2 lesion at w096, given pre-treatment and w048 data. To estimate the number of NET2 lesions in a mask generated by our model, we apply a morphological opening operation using a $1\times5\times5$ kernel to refine lesion boundaries, and then perform connected component analysis (CCA). For the ground truth count, we perform an identical CCA on the ground truth mask. Performance is measured using average precision (AP) and the area under the receiver operating characteristic curve (AUC). As a baseline, we include a Monte Carlo classifier that samples binary outcomes based on the observed frequency of positive cases at w096 in each treatment arm of the training set. Table~\ref{tab:activity_prediction}, shows that our proposed model outperforms this baseline across each treatment group.

\begin{table}[!htbp]
\centering
\caption{Binary outcome prediction performance of the proposed model compared to a baseline built from population-level statistics. MC baseline metrics reported based on an average of 100 runs. Results reported on the test set}
\label{tab:activity_prediction}
\begin{tabular}{llccccccc}
    \toprule
     & \textbf{Metric} & \textbf{Placebo} & \textbf{NE} & \textbf{LE} & \textbf{MiE} & \textbf{MoE} & \textbf{HE} \\
    \midrule
    \multirow{2}{*}{Proposed Method} 
    & \textbf{AP}~$\uparrow$    & \textbf{0.783} & \textbf{0.735} & \textbf{0.623} & \textbf{0.647} & \textbf{0.436} & \textbf{0.083} \\
    & \textbf{AUC}~$\uparrow$   & \textbf{0.826} & \textbf{0.662} & \textbf{0.536} & \textbf{0.705} & \textbf{0.607} & \textbf{0.868} \\
    \midrule
    \multirow{2}{*}{MC Baseline} 
    & \textbf{AP}~$\uparrow$  & 0.513   & 0.577   & 0.517   & 0.361   & 0.342   & 0.019 \\
    & \textbf{AUC}~$\uparrow$ & 0.495   & 0.505   & 0.487   & 0.507   & 0.489   & 0.509 \\
    \bottomrule
\end{tabular}
\end{table}

\paragraph{NET2 Count Prediction:} To further assess the models' performance, we compute the mean squared error (MSE) of the logarithm of the predicted NET2 lesion count, following~\cite{durso-finley22a}. Applying the logarithm helps reduce the influence of outliers, which is critical given the long-tailed distribution of NET2 lesion counts in MS. As a baseline, we use a predictor that assigns the average lesion count of each treatment group in the training set. As shown in Table~\ref{tab:lesion_counts}, our model consistently outperforms this baseline across all treatment groups.

\begin{table}[!htbp]
    \centering
    \renewcommand{\arraystretch}{1.2}
    \caption{MSE of log NET2 count against an average baseline on the test set.} 
    \begin{tabular}{lcccccc}
        \hline
        & \textbf{Placebo}~$\downarrow$ & \textbf{NE}~$\downarrow$ & \textbf{LE}~$\downarrow$ & \textbf{MiE}~$\downarrow$ & \textbf{MoE}~$\downarrow$ & \textbf{HE}~$\downarrow$ \\
        \hline
        Proposed Method    & \textbf{1.538} & \textbf{1.158} & \textbf{1.093} & \textbf{0.842} & \textbf{0.531} & \textbf{0.006} \\ \hline
        Avg. Baseline & 1.775 & 1.620 & 1.780 & 0.991 & 0.697 & 0.010 \\
        \hline
    \end{tabular}
    \label{tab:lesion_counts}
\end{table}

\subsection{Exploring Counterfactual Predictions}

\begin{figure}[!!htbp]
    \centering
    \includegraphics[width=\linewidth]{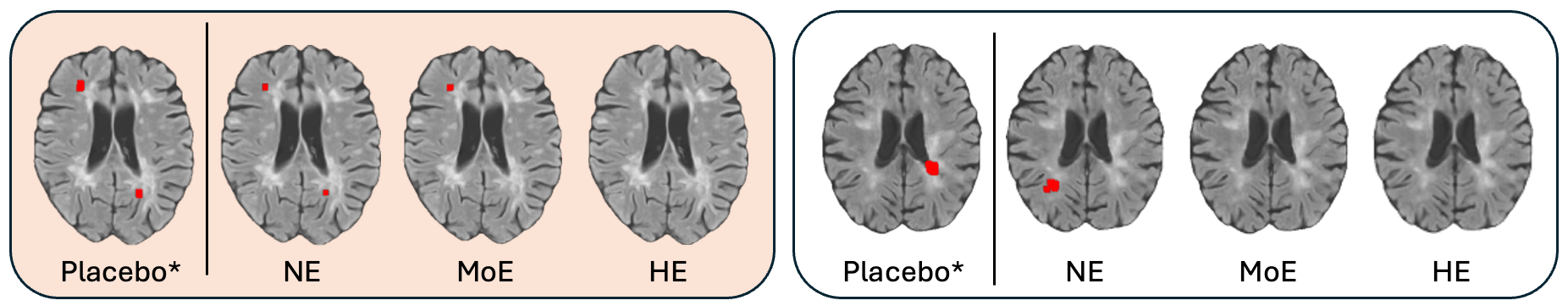}
    \caption{Counterfactual predictions of future NET2 lesion masks for a single patient under treatments of varying efficacies, which increase from left to right. The model, conditioned only on pre-treatment data, reflects reduced future NET2 under more effective therapies, illustrating its potential for individualized treatment effect estimation.}
    \label{fig:ite-qualitative}
\end{figure}

Last, we explore the proposed model's potential to perform counterfactual image generation, specifically of a patient's future NET2 lesion mask, under different treatment conditions. We re-train the model using only pre-treatment images (w000) as conditioning information, and make predictions for a given patient under various treatments at w096. A few qualitative results are shown in Figure~\ref{fig:ite-qualitative}. These results illustrate how treatments with different efficacies show reductions in NET2 lesions in the associated predicted images. While a detailed quantitative analysis of treatment effects is left for future work, these results highlight the promise of generative models in predicting future treatment-specific counterfactual medical images. 

\section{Discussion}

Predicting the future evolution of NET2 lesions in Multiple Sclerosis (MS) is a profoundly complex task, made challenging by both biological and technical factors. From a biological standpoint, the progression of MS is known to be highly heterogeneous~\cite{lucchinetti2000heterogeneity}. Lesion development patterns vary significantly not only between population subgroups but also between individuals. Since NET2 lesion progression can follow multiple plausible trajectories, the observed outcome represents only one of several biologically reasonable possibilities. Consequently, standard metrics like accuracy, Dice score, or MSE may penalize valid predictions that deviate from this single outcome, overlooking clinically meaningful variation. This makes it difficult not only to model the disease but also to justify reasonable metrics for evaluating performance.

In addition to biological complexity, technical challenges inherent to medical imaging further complicate predictive modeling. MRI data, in particular, suffers from substantial variability due to differences in acquisition protocols, scanner hardware, spatial resolution, and contrast characteristics. While preprocessing steps like registration, normalization, and artifact correction are employed to mitigate these effects, such pipelines are imperfect and can introduce residual noise and inconsistencies that divert model capacity from true disease-related patterns. This is of course in addition to label noise and inter-rater variability in manual segmentation of NET2 lesions which can further impact the reliability of ground truth annotations.

It is also important to contextualize the Monte Carlo baselines used in our evaluations. While they may seem simplistic, they reflect real-world clinical heuristics: in the absence of individualized models, population-level statistics often form the basis for informing patients about prognosis. Thus, these baselines are not without value. However, as machine learning approaches mature, we anticipate that our models, baselines, and relevant metrics will evolve in sophistication and clinical relevance in personalized decision making.

Future research directions can, first, incorporate richer sources of patient-specific data as conditioning signals, including demographic, genetic, and clinical metrics for greater predictive accuracy. Second, a more formal, quantitative analysis of causal treatment effects at the individual level would extend the current qualitative findings presented in this work. Lastly, integrating MS-specific prognostic models, like~\cite{durso-finley22a}, into the generation process could lead to improved performance by providing meaningful and structured prior knowledge~\cite{BRLP}.

\section{Conclusion}

In this work, we present the first treatment-aware conditional diffusion framework that forecasts the evolution of new and enlarging T2 (NET2) lesions in MS. Our approach employs a pseudo-2D representation of 3D MRI volumes, enabling high-resolution, voxel-level predictions without reliance on external anatomical priors. Comprehensive evaluations demonstrate that the model effectively predicts disease activity and provides interpretable, patient-specific insights into treatment responses. Our findings highlight the potential of generative AI to advance personalized medicine by allowing clinicians to visualize possible treatment responses, thereby moving towards AI-powered decision support for complex neurological disorders.

\begin{credits}
\subsubsection{\ackname} This investigation was supported by the International Progressive Multiple Sclerosis Alliance (PA-1412-02420) and the companies who generously provided the data: Biogen, Roche/Genentech, and Teva; the MS Society of Canada; the Natural Sciences and Engineering Research Council of Canada; Fonds de Recherche du Quebec: Nature et Technologies; the Canadian Institute for Advanced Research (CIFAR) Artificial Intelligence Chairs program; Google Research; Calcul Quebec; the Digital Research Alliance of Canada; and Mila - Quebec AI Institute.

\subsubsection{\discintname} The authors have no competing interests to declare.
\end{credits}
%
%
%
%
\newpage
\bibliographystyle{splncs04}
\bibliography{main}

\begin{thebibliography}{10}
\providecommand{\url}[1]{\texttt{#1}}
\providecommand{\urlprefix}{URL }
\providecommand{\doi}[1]{https://doi.org/#1}

\bibitem{Billot_2023}
Billot, B., Greve, D.N., Puonti, O., Thielscher, A., Van~Leemput, K., Fischl, B., Dalca, A.V., Iglesias, J.E.: Synthseg: Segmentation of brain mri scans of any contrast and resolution without retraining. Medical Image Analysis  \textbf{86},  102789 (May 2023). \doi{10.1016/j.media.2023.102789}, \url{http://dx.doi.org/10.1016/j.media.2023.102789}

\bibitem{dhariwal2021diffusionmodelsbeatgans}
Dhariwal, P., Nichol, A.: Diffusion models beat gans on image synthesis. Eprint \href{https://arxiv.org/abs/2105.05233}{arXiv:2105.05233} (2021)

\bibitem{Doyle2017LesionDS}
Doyle, A., Elliott, C., Karimaghaloo, Z., Subbanna, N.K., Arnold, D.L., Arbel, T.: Lesion detection, segmentation and prediction in multiple sclerosis clinical trials. In: BrainLes@MICCAI (2017), \url{https://api.semanticscholar.org/CorpusID:3440295}

\bibitem{dursofinley2023}
Durso-Finley, J., Falet, J.P., Mehta, R., Arnold, D.L., Pawlowski, N., Arbel, T.: Improving image-based precision medicine with uncertainty-aware causal models (2023), \url{https://arxiv.org/abs/2305.03829}

\bibitem{durso-finley22a}
Durso-Finley, J., Falet, J.P., Nichyporuk, B., Douglas, A., Arbel, T.: Personalized prediction of future lesion activity and treatment effect in multiple sclerosis from baseline mri. In: Konukoglu, E., Menze, B., Venkataraman, A., Baumgartner, C., Dou, Q., Albarqouni, S. (eds.) Proceedings of The 5th International Conference on Medical Imaging with Deep Learning. Proceedings of Machine Learning Research, vol.~172, pp. 387--406. PMLR (06--08 Jul 2022), \url{https://proceedings.mlr.press/v172/durso-finley22a.html}

\bibitem{favero2025conditionaldiffusionmodelsmedical}
Favero, G.M., Saremi, P., Kaczmarek, E., Nichyporuk, B., Arbel, T.: Conditional diffusion models are medical image classifiers that provide explainability and uncertainty for free (2025), \url{https://arxiv.org/abs/2502.03687}

\bibitem{Friedrich_2024}
Friedrich, P., Wolleb, J., Bieder, F., Durrer, A., Cattin, P.C.: WDM: 3D Wavelet Diffusion Models for High-Resolution Medical Image Synthesis, p. 11–21. Springer Nature Switzerland (Oct 2024). \doi{10.1007/978-3-031-72744-3_2}, \url{http://dx.doi.org/10.1007/978-3-031-72744-3_2}

\bibitem{MSLesionCoeff}
Ge, T., Müller-Lenke, N., Bendfeldt, K., Nichols, T., Johnson, T.: Analysis of multiple sclerosis lesions via spatially varying coefficients. The Annals of Applied Statistics  \textbf{8} (06 2014). \doi{10.1214/14-AOAS718}

\bibitem{Gold2012}
Gold, R., Kappos, L., Arnold, D.L., Bar-Or, A., Giovannoni, G., Selmaj, K., Tornatore, C., Sweetser, M.T., Yang, M., Sheikh, S.I., Dawson, K.T., Investigators, D.S.: Placebo-controlled phase 3 study of oral {BG-12} for relapsing multiple sclerosis. The New England Journal of Medicine  \textbf{367}(12),  1098--1107 (2012). \doi{10.1056/NEJMoa1114287}, \url{https://doi.org/10.1056/NEJMoa1114287}

\bibitem{hang2024efficientdiffusiontrainingminsnr}
Hang, T., Gu, S., Li, C., Bao, J., Chen, D., Hu, H., Geng, X., Guo, B.: Efficient diffusion training via min-snr weighting strategy. Eprint \href{https://arxiv.org/abs/2303.09556}{arXiv:2303.09556} (2024)

\bibitem{Hauser2017}
Hauser, S.L., Bar-Or, A., Comi, G., Giovannoni, G., Hartung, H.P., Hemmer, B., Lublin, F., Montalban, X., Rammohan, K.W., Selmaj, K., Traboulsee, A., Wolinsky, J.S., Arnold, D.L., Klingelschmitt, G., Masterman, D., Fontoura, P., Belachew, S., Chin, P., Mairon, N., Garren, H., I, O., Investigators, O.I.C.: Ocrelizumab versus interferon beta-1a in relapsing multiple sclerosis. The New England Journal of Medicine  \textbf{376}(3),  221--234 (2017). \doi{10.1056/NEJMoa1601277}, \url{https://doi.org/10.1056/NEJMoa1601277}

\bibitem{ho2022classifierfreediffusionguidance}
Ho, J., Salimans, T.: Classifier-free diffusion guidance. Eprint \href{https://arxiv.org/abs/2207.12598}{arXiv:2207.12598} (2022)

\bibitem{Hoogeboom:arXiv:2023:simpleDiffusion}
Hoogeboom, E., Heek, J., Salimans, T.: Simple diffusion: End-to-end diffusion for high resolution images. Eprint \href{https://arxiv.org/abs/2301.11093}{arXiv:2301.11093} (2023)

\bibitem{Khader2023}
Khader, F., Müller-Franzes, G., Tayebi~Arasteh, S., et~al.: Denoising diffusion probabilistic models for 3d medical image generation. Scientific Reports  \textbf{13}, ~7303 (2023). \doi{10.1038/s41598-023-34341-2}, \url{https://doi.org/10.1038/s41598-023-34341-2}

\bibitem{kim2024adaptive}
Kim, J., Park, H.: Adaptive latent diffusion model for 3d medical image to image translation: Multi-modal magnetic resonance imaging study. In: Proceedings of the IEEE/CVF Winter Conference on Applications of Computer Vision. pp. 7604--7613 (2024)

\bibitem{kingma2023variationaldiffusionmodels}
Kingma, D.P., Salimans, T., Poole, B., Ho, J.: Variational diffusion models. Eprint \href{https://arxiv.org/abs/2107.00630}{arXiv:2107.00630} (2023)

\bibitem{MSLesionCoeff2}
Lee, M.A., Smith, S., Palace, J., Narayanan, S., Silver, N., Minicucci, L., Filippi, M., Miller, D.H., Arnold, D.L., Matthews, P.M.: Spatial mapping of t2 and gadolinium-enhancing t1 lesion volumes in multiple sclerosis: evidence for distinct mechanisms of lesion genesis? Brain  \textbf{122}(7),  1261--1270 (07 1999). \doi{10.1093/brain/122.7.1261}, \url{https://doi.org/10.1093/brain/122.7.1261}

\bibitem{lucchinetti2000heterogeneity}
Lucchinetti, C., Br{\"u}ck, W., Parisi, J., Scheithauer, B., Rodriguez, M., Lassmann, H.: Heterogeneity of multiple sclerosis lesions: implications for the pathogenesis of demyelination. Annals of Neurology  \textbf{47}(6),  707--717 (Jun 2000). \doi{10.1002/1531-8249(200006)47:6<707::AID-ANA3>3.0.CO;2-Q}

\bibitem{BRLP}
Puglisi, L., Alexander, D.C., Ravì, D.: Enhancing spatiotemporal disease progression models via latent diffusion and prior knowledge (2024), \url{https://arxiv.org/abs/2405.03328}

\bibitem{Rachmadi2025}
Rachmadi, M.F., Valdés-Hernández, M.C., Makin, S., et~al.: Prediction of white matter hyperintensities evolution one-year post-stroke from a single-point brain mri and stroke lesions information. Scientific Reports  \textbf{15}, ~1208 (2025). \doi{10.1038/s41598-024-83128-6}, \url{https://doi.org/10.1038/s41598-024-83128-6}

\bibitem{ronneberger2015unet}
Ronneberger, O., Fischer, P., Brox, T.: U-net: Convolutional networks for biomedical image segmentation. Medical Image Computing and Computer-Assisted Intervention (MICCAI)  \textbf{9351},  234--241 (2015). \doi{10.1007/978-3-319-24574-4_28}

\bibitem{Sepahvand2020CNNDO}
Sepahvand, N.M., Arnold, D.L., Arbel, T.: Cnn detection of new and enlarging multiple sclerosis lesions from longitudinal mri using subtraction images. 2020 IEEE 17th International Symposium on Biomedical Imaging (ISBI) pp. 127--130 (2020), \url{https://api.semanticscholar.org/CorpusID:218895124}

\bibitem{Vollmer2014}
Vollmer, T.L., Sorensen, P.S., Selmaj, K., Zipp, F., Havrdova, E., Cohen, J.A., Sasson, N., Gilgun-Sherki, Y., Arnold, D.L., Group, B.S.: A randomized placebo-controlled phase {III} trial of oral laquinimod for multiple sclerosis. Journal of Neurology  \textbf{261}(4),  773--783 (2014). \doi{10.1007/s00415-014-7264-4}, \url{https://doi.org/10.1007/s00415-014-7264-4}

\bibitem{wang20243d}
Wang, H., Liu, Z., Sun, K., Wang, X., Shen, D., Cui, Z.: 3d meddiffusion: A 3d medical diffusion model for controllable and high-quality medical image generation. arXiv preprint arXiv:2412.13059  (2024)

\bibitem{sadm}
Yoon, J.S., Zhang, C., Suk, H.I., Guo, J., Li, X.: SADM: Sequence-Aware Diffusion Model for Longitudinal Medical Image Generation, p. 388–400. Springer Nature Switzerland (2023). \doi{10.1007/978-3-031-34048-2_30}, \url{http://dx.doi.org/10.1007/978-3-031-34048-2_30}

\bibitem{zhang2023addingconditionalcontroltexttoimage}
Zhang, L., Rao, A., Agrawala, M.: Adding conditional control to text-to-image diffusion models. Eprint \href{https://arxiv.org/abs/2302.05543}{arXiv:2302.05543} (2023)

\end{thebibliography}
\end{document}